\documentclass{article}

\usepackage{amssymb,amsmath}
\usepackage{graphicx}
\usepackage{subcaption}
\usepackage[authoryear]{natbib}
\usepackage{authblk}

\def\Ttheta{\boldsymbol\theta}

\def\TC{\mathbf C}

\def\TK{\mathbf K}

\def\TY{\mathbf Y}

\def\TW{\mathbf W}

\def\TX{\mathbf X}

\begin{document}

\title{Fully reversible neural networks for large-scale 3D seismic horizon tracking}

\author[1]{Bas Peters}
\author[2]{Eldad Haber}

\affil[1]{Computational Geosciences Inc., Vancouver, Canada}
\affil[2]{University of British Columbia, Vancouver, Canada}
\date{January 15, 2020}
\maketitle

\section{Introduction}
The primary aim of this work is to address memory limitations of 3D convolutional neural networks for seismic interpretation. Horizon tracking based on a few label picks (seed points) is an important task in the structural interpretation of seismic images and 3D volumes. Because manual interpretation is time-consuming, researchers have worked on various methods to accelerate the horizon tracking from a few manual picks. See \cite{doi:10.1190/geo2017-0830.1} for a comparison of a few recent state-of-the-art methods not based on learning. Seismic interpretation was an early adopter of neural network technology, e.g., \citep{A226265}, using shallow networks acting on 1D traces or 2D seismic images. More recently, researchers employed networks such as auto-encoders and U-nets \citep{Ronneberger2015}, to train a map from 2D seismic image to a 2D horizon image, e.g., \citep{wu2018deep,di2018developing,doi:10.1190/INT-2018-0225.1,doi:10.1190/tle38070534.1}.

In order to exploit 3D structure, we would like to train neural networks with 3D convolutional kernels. While such networks exist and are employed for various seismic interpretation tasks, the largest 3D inputs for fault segmentation \citep{doi:10.1190/geo2018-0646.1,doi:10.1190/segam2019-3216307.1}, salt segmentation \citep{doi:10.1190/segam2018-2997304.1}, and seismic image processing \citep{doi:10.1190/INT-2018-0224.1} seem to be $128 \times 128 \times 128$ pixels. This shows that the spatial range is small compared to the $1088 \times 2816$ pixel 2D images used by \cite{doi:10.1190/INT-2018-0225.1}. The main roadblock to increase the input size of the 3D data is the limited memory on graphical processing units (GPUs), while the requirements for the network weights and states (activations) may be large. Particularly, the network states are required to be in memory during the computation of a gradient of the loss function, if we employ standard reverse-mode automatic differentiation. The memory demands then grow with data size, number of layers in the network, and the number of channels in each layer.

For the first time, we use \emph{fully reversible networks} for the interpretation of 3D seismic imaging results. These networks have memory requirements that are independent of the depth of the network; they require storage of the network states for three layers only in order to compute a gradient of the loss function. Therefore, fully reversible networks need much less memory compared to non-reversible networks (including U-nets). Instead, we allocate the saved memory toward using larger input data. In this work, we demonstrate that \emph{a}) fully reversible networks based on hyperbolic differential equations are suitable for detecting and tracking horizons in seismic image volumes; \emph{b}) examples on real data from the North Sea where our input for the neural network is about $15\times$ larger than recently published 3D convolutional neural networks for seismic interpretation.

Because this is the first work that employs reversible neural networks for seismic interpretation in 3D, there are still many questions left unanswered. We are concerned with computational issues and the feasibility of our method. While we present `good' results on real data, a full comparison with networks based on 2D slices in terms of computational cost and prediction quality is out of our scope.

\section{Problem formulation for 3D seismic horizon tracking}
Our aim is to track horizons through a 3D seismic volume based a few manual horizon picks (seed points). There is a single dataset and generalization to other seismic data is not the goal. We take a non-linear regression approach and minimize the non-linear least-squares problem
\begin{equation}\label{loss}
\phi(\Ttheta,\TX,\TC) = \frac{1}{2} \sum_{i \in \Omega} \| (\TY_n(\TX,\Ttheta)_i - \TC_i \|_2^2.
\end{equation}
For one example, the input for the neural network, $\TX \in \mathbb{R}^{n_x \times n_y \times n_z \times n_c}$, is the 3D seismic imaging result with $n_c$ channels, such as imaging from near, mid, and far offset recordings. The collection of network parameters, $\Ttheta$, includes convolutional kernels. Our network contains just convolutional kernels, so we can change the size of the input at any time. The output of the neural network at the final layer (layer $n$), $\TY_n$, depends on the input data and network parameters. For the construction of the labels, $\TC \in \mathbb{R}^{n_x \times n_y \times n_z \times n_c}$, we follow a similar procedure as in \cite{doi:10.1190/INT-2018-0225.1}. For each training and validation horizon pick, we construct the label as $1$ minus the distance from the horizon pick in the vertical direction only. We clip the maximum at some small distance from the horizon pick so that there are only positive numbers in the label. The label tensor is thus a sparsely populated cube, where each fiber (`column' in the vertical direction) at a horizon point is as shown in Figure \ref{fig:pred_plus_label}. The indices of the label cube where we have the horizon picks and therefore the labels in the full vertical extent at that spatial location, are collected in the set $\Omega$. The partial loss function measures the misfit at the known label locations only, and computes $\nabla_{\Ttheta} \phi(\Ttheta,\TX,\TC)$ based on those locations, see \cite{doi:10.1190/INT-2018-0225.1,doi:10.1190/tle38070534.1} for why training a network using a partial loss function is equivalent to solving geophysical inverse problems such as seismic full waveform inversion.

\section{Fully reversible hyperbolic networks with factorized layers}
This section outlines our primary contribution. The layers of the neural network in \eqref{loss} are a fully reversible hyperbolic neural network \citep{lensink2019fully} to map the data volume to horizon cube output. A reversible hyperbolic network is based on a conservative leapfrog discretization of a non-linear telegraph equation \citep{Chang2017Reversible} as follows:
\begin{align}\label{network}
\TY_1 & = \TX, \: \TY_2 = \TX \nonumber \\
\TY_{j} &= 2 \TW_{j-1}\TY_{j-1} -  \TW_{j-2} \TY_{j-2} -  h^2 \TK^\top_{j-1} f( \TK_{j-1} \TW_{j-1} \TY_{j-1}), \: j=3,\cdots,n \nonumber  \\
\end{align}
where the first two layers indicate the initial conditions that are equal to the input data $\TX$. The non-linear pointwise activation function is the ReLU in this work, $f(x) = \operatorname{max}(0,x)$. The `time-step' $h$ determines the stability of the network and needs to be chosen sufficiently small. The number of channels changes simultaneously with resolution (coarsening, pooling). The linear operator $\TW$ is an orthogonal Haar wavelet transform if we change resolution and channels at a given layer; otherwise $\TW$ is the identity map. Convolutional kernels $\Ttheta$ at layer $j$ are included in $\TK_j$. To reduce the number of channels and increase the resolution, we use the inverse Haar transform, which is known in closed form. The combination of the leapfrog discretization of the telegraph equation that models signal propagation, and an invertible transform to change channels/resolution, allows for full reversal of the network via \citep{lensink2019fully}
\begin{equation}\label{rev_prop}
\TY_j = \TW_j^{-1} \bigg[ 2 \TW_{j+1} \TY_{j+1} -  h^2 \TK_{j+1}^\top  f ( \TK_{j+1} \TW_{j+1} \TY_{j+1} ) - \TY_{j+2} \bigg], \: j=n-2,\cdots,3.
\end{equation}
Therefore, we can re-compute the network states $\TY_j$ while back-propagating to compute the gradient of $\phi(\Ttheta,\TX,\TC)$ w.r.t. $\Ttheta$. This avoids the need to store all network states for every layer, which is the limiting factor to employ deep neural networks on large 3D/4D computational domains. Using the fully reversible network, we need to store the states for three layers only. This situation is equivalent to how we need to store the wavefield for every time step to compute a gradient for full-waveform inversion. Stability of the forward and reverse propagation is always easy to verify numerically. Note that by construction of the network \eqref{rev_prop}, we do not need to invert/reverse any activation functions. The fully reversible construction changes a linear memory growth with network depth, into a constant. In the next section, we use the saved memory to increase the data input size to be able to learn from larger scale structure.

The various U-nets, auto-encoders, as well as our proposed network so far operate on multiple resolutions. Increasing/decreasing resolution via invertible transform, such as orthogonal Haar wavelets, increases the number of channels by a factor $8\times$ in 3D. Starting with three input channels and coarsening three times thus results in $3 \times 8^3 = 1536$ channels. This would standardly require the storage of $1536^2$ convolutional kernels for each layer at the coarsest resolution. To avoid rapidly increasing memory demands when coarsening inside a fully reversible neural network, we use the trick from \cite{peters2019symmetric} to parameterize the weights at a layer as a symmetric block-low-rank structure. Each block of the block-matrix $\TK$ contains a convolution matrix representation corresponding to a kernel $\theta$. To maintain a reversible structure of the network but at the same time reduce the number of convolutional kernels in a given layer, we set up a block-low-rank version of $\TK^\top f( \TK \TY)$ as follows:
   
\begin{equation}\label{BLRLayer}
\begin{bmatrix}
K(\Ttheta^{1,1})^\top & K(\Ttheta^{2,1})^\top\\
K(\Ttheta^{1,2})^\top & K(\Ttheta^{2,2})^\top\\
K(\Ttheta^{1,3})^\top & K(\Ttheta^{2,3})^\top\\
K(\Ttheta^{1,4})^\top & K(\Ttheta^{2,4})^\top
\end{bmatrix}
f \bigg(
\begin{bmatrix}
K(\Ttheta^{1,1}) & K(\Ttheta^{1,2}) & K(\Ttheta^{1,3}) & K(\Ttheta^{1,4})\\
K(\Ttheta^{2,1}) & K(\Ttheta^{2,2}) & K(\Ttheta^{2,3}) & K(\Ttheta^{2,4})\\
\end{bmatrix}
\begin{bmatrix}
Y^1 \\
Y^2 \\
Y^3 \\
Y^4 \\
\end{bmatrix} \bigg).
\end{equation}
This is an example where we have four input and output channels using only eight kernels instead of the usual $16$. The block-matrix notation is for linear algebraic derivations, but the computational implementation can still be in 5D tensor format. In the example in the following section we obtain good predictions when using just $32 \times 1536$ kernels in $\TK$ instead of the usual $1536 \times 1536$; a $48\times$ reduction.

\section{North Sea data example}
An industrial partner provided a section of size $1152 \times 1152 \times 288$ of seismic imaging from the North Sea that we split into six examples of size $192 \times 192 \times 288$, as well as manual horizon picks. For training labels, we randomly select eight horizon picks in each 3D sub-volume, and the remaining picks are for validation. Table \ref{network_design} lists the network details. The memory for three network layer states $\TY_j$ is $\approx382$ megabyte, which would be $\approx2675$ in case of non-reversible networks that need to store all states to compute a gradient of the loss. The block-low-rank layers \eqref{BLRLayer} need $\approx 40$ megabyte for network weights instead $\approx 1807$ for regular layers. These memory savings allow for input data that is about $15\times$ larger than the $128^3$ used in some recently published geophysical work with 3D convolutional networks \citep{doi:10.1190/geo2018-0646.1,doi:10.1190/segam2019-3216307.1,doi:10.1190/segam2018-2997304.1,doi:10.1190/INT-2018-0224.1}. Therefore, fully reversible networks bring us a step closer to working with the length scales that are feasible in case of 2D convolutional neural networks.

\begin{table}[!htb]
\centering
\begin{tabular}{r|r|c|c|c}
Layer \# &  Feature size & \# of channels & \# of kernels, shape of $\TK_j$ & kernel size  \\
\hline
3-8   & $24 \times 24 \times 36$ & $1536$ & $32 \times 1536$   & $3\times3 \times 3$\\
9-14   & $48 \times 48 \times 72$ & $192$ & $24 \times 192$ & $3\times3 \times 3$\\	
15-19   & $96 \times 96 \times 144$ & $24$ & $8 \times 24$  & $3\times3 \times 3$\\	
20-24   & $192 \times 192 \times 288$ & $3$ & $3 \times 3 $  & $3\times3 \times 3$\\	
\end{tabular}
\caption{Design of our convolutional, fully reversible hyperbolic network with symmetric block-low-rank layers. We start at a coarser resolution, generated by applying the 3D Haar transform three times to the input data volumes of size $192 \times 192 \times 288 \times 3$.}\label{network_design}
\end{table}

\begin{figure}[!htb]
   \centering
   \includegraphics[width=0.7\textwidth]{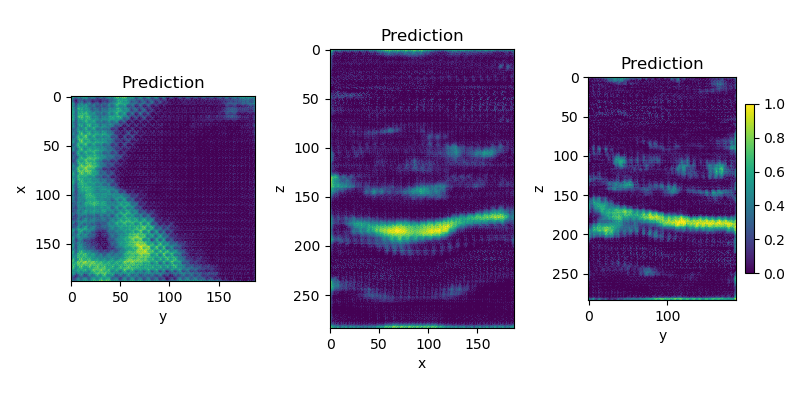}
   \caption{Three slices through one of the 3D predicted horizon volumes.}
   \label{fig:prediction}
 \end{figure}

\begin{figure}[!htb]
 	\centering
 	\begin{subfigure}[b]{0.49\textwidth}
 		\includegraphics[width=\textwidth]{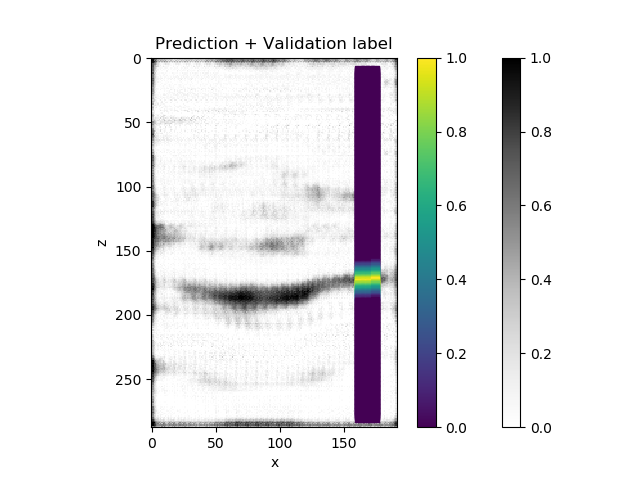}
 		\caption{}
 		\label{fig:Figure1a}
 	\end{subfigure}
 	\begin{subfigure}[b]{0.49\textwidth}
 		\includegraphics[width=\textwidth]{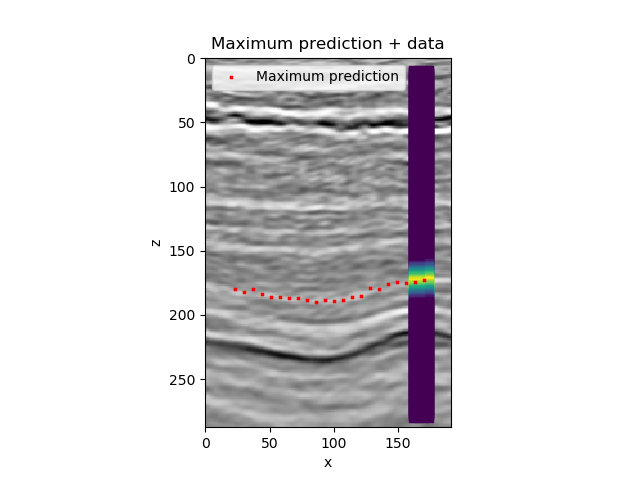}
 		\caption{}
 		\label{fig:Figure1b}
 	\end{subfigure}
 	\caption{a) A slice through the 3D predicted horizon volume, and the validation labels. b) A slice through the seismic image volume, the most likely predicted horizon location, and the validation labels.}
\label{fig:pred_plus_label}
 \end{figure}

An example three cross-sections of the final prediction is shown in Figure \ref{fig:prediction}. For the x-z slice, we also show the validation label overlaid on the prediction (Figure \ref{fig:pred_plus_label}a), and the highest predicted value for each x-y location in Figure  \ref{fig:pred_plus_label}b.

\section{Conclusions}
We introduced fully reversible convolutional neural networks to seismic horizon tracking on large 3D inputs. These recently proposed networks have a memory requirement that is independent of network depth, and there is no need to store the network states for every layer in the network to compute a gradient of the loss. The huge memory savings compared to previously proposed 3D convolutional neural networks enable us to increase the input size of the 3D data, brining the length scales of the input a step closer to what is possible using 2D inputs. A field data example validated that the proposed network is suitable for seismic interpretation and that the input data size for the network can be increased by an order of magnitude using fully reversible neural networks, if combined with a network weight parameterization strategy such as a block-low-rank layer structure.

\bibliography{biblio.bib}

\begin{thebibliography}{15}
\providecommand{\natexlab}[1]{#1}

\bibitem[Chang et~al., 2018]{Chang2017Reversible}
Chang, B., Meng, L., Haber, E., Ruthotto, L., Begert, D. and Holtham, E. [2018]
  Reversible Architectures for Arbitrarily Deep Residual Neural Networks.
\newblock In: \emph{AAAI Conference on AI}.

\bibitem[Di, 2018]{di2018developing}
Di, H. [2018] Developing a seismic pattern interpretation network (SpiNet) for
  automated seismic interpretation.
\newblock \emph{arXiv preprint arXiv:1810.08517}.

\bibitem[Harrigan et~al., 1992]{A226265}
Harrigan, E., Kroh, J.R., Sandham, W.A. and Durrani, T.S. [1992] Seismic
  horizon picking using an artificial neural network.
\newblock In: \emph{[Proceedings] ICASSP-92: 1992 IEEE International Conference
  on Acoustics, Speech, and Signal Processing}, ~3. 105--108 vol.3.

\bibitem[Lensink et~al., 2019]{lensink2019fully}
Lensink, K., Haber, E. and Peters, B. [2019] Fully Hyperbolic Convolutional
  Neural Networks.
\newblock \emph{arXiv preprint arXiv:1905.10484}.

\bibitem[Peters et~al., 2019{\natexlab{a}}]{doi:10.1190/INT-2018-0225.1}
Peters, B., Granek, J. and Haber, E. [2019{\natexlab{a}}] Multiresolution
  neural networks for tracking seismic horizons from few training images.
\newblock \emph{Interpretation}, \textbf{7}(3), SE201--SE213.

\bibitem[Peters et~al., 2019{\natexlab{b}}]{doi:10.1190/tle38070534.1}
Peters, B., Haber, E. and Granek, J. [2019{\natexlab{b}}] Neural networks for
  geophysicists and their application to seismic data interpretation.
\newblock \emph{The Leading Edge}, \textbf{38}(7), 534--540.

\bibitem[Peters et~al., 2019{\natexlab{c}}]{peters2019symmetric}
Peters, B., Haber, E. and Lensink, K. [2019{\natexlab{c}}] Symmetric
  block-low-rank layers for fully reversible multilevel neural networks.
\newblock \emph{arXiv preprint arXiv:1912.12137}.

\bibitem[Ronneberger et~al., 2015]{Ronneberger2015}
Ronneberger, O., Fischer, P. and Brox, T. [2015] U-Net: Convolutional Networks
  for Biomedical Image Segmentation.
\newblock In: Navab, N., Hornegger, J., Wells, W.M. and Frangi, A.F. (Eds.)
  \emph{Medical Image Computing and Computer-Assisted Intervention -- MICCAI
  2015}. Springer International Publishing, Cham, 234--241.

\bibitem[Shi et~al., 2018]{doi:10.1190/segam2018-2997304.1}
Shi, Y., Wu, X. and Fomel, S. [2018] Automatic salt-body classification using
  deep-convolutional neural network.
\newblock In: \emph{SEG Technical Program Expanded Abstracts 2018}. 1971--1975.

\bibitem[Veezhinathan et~al., 1993]{Veezhinathan1993}
Veezhinathan, J., Kemp, F. and Threet, J. [1993] A Hybrid of Neural Net and
  Branch and Bound Techniques for Seismic Horizon Tracking.
\newblock In: \emph{Proceedings of the 1993 ACM/SIGAPP Symposium on Applied
  Computing: States of the Art and Practice}, SAC '93. ACM, New York, NY, USA,
  173--178.

\bibitem[Wang and Nealon, 2019]{doi:10.1190/INT-2018-0224.1}
Wang, E. and Nealon, J. [2019] Applying machine learning to 3D seismic image
  denoising and enhancement.
\newblock \emph{Interpretation}, \textbf{7}(3), SE131--SE139.

\bibitem[Wu and Zhang, 2018]{wu2018deep}
Wu, H. and Zhang, B. [2018] A deep convolutional encoder-decoder neural network
  in assisting seismic horizon tracking.
\newblock \emph{arXiv preprint arXiv:1804.06814}.

\bibitem[Wu and Fomel, 2018]{doi:10.1190/geo2017-0830.1}
Wu, X. and Fomel, S. [2018] Least-squares horizons with local slopes and
  multigrid correlations.
\newblock \emph{GEOPHYSICS}, \textbf{83}(4), IM29--IM40.

\bibitem[Wu et~al., 2019]{doi:10.1190/geo2018-0646.1}
Wu, X., Liang, L., Shi, Y. and Fomel, S. [2019] FaultSeg3D: Using synthetic
  data sets to train an end-to-end convolutional neural network for 3D seismic
  fault segmentation.
\newblock \emph{GEOPHYSICS}, \textbf{84}(3), IM35--IM45.

\bibitem[Zhao, 2019]{doi:10.1190/segam2019-3216307.1}
Zhao, T. [2019] 3D convolutional neural networks for efficient fault detection
  and orientation estimation.
\newblock In: \emph{SEG Technical Program Expanded Abstracts 2019}. 2418--2422.

\end{thebibliography}
\end{document}